\documentclass[12pt]{iopart}
\usepackage{iopams}  
\usepackage{epsfig} %eps figures package
\usepackage{color}

\begin{document}
\bibliographystyle{aip}

\title[Organic Magnetoresistance]{Large magnetoresistance at room-temperature in small molecular weight organic semiconductor sandwich devices}

\author{\"O. Mermer\dag\, G. Veeraraghavan\ddag\, T. L. Francis\ddag\, M. Wohlgenannt\dag\ \footnote[3]{To
whom correspondence should be addressed (markus-wohlgenannt@uiowa.edu)}
}

\address{\dag\ Department of Physics and Astronomy, The University of Iowa, Iowa City, IA 52242-1479, USA}
\address{\ddag\ Department of Electrical and Computer Engineering, The University of Iowa, Iowa City, IA 52242-1595, USA}

\begin{abstract}
We present an extensive study of a large, room temperature negative magnetoresistance (MR) effect in tris-(8-hydroxyquinoline) aluminum sandwich devices in weak magnetic fields. The effect is similar to that previously discovered in polymer devices. We characterize this effect and discuss its dependence on field direction, voltage, temperature, film thickness, and electrode materials. The MR effect reaches almost 10\% at fields of approximately 10 mT at room temperature. The effect shows only a weak temperature dependence and is independent of the sign and direction of the magnetic field. Measuring the devices' current-voltage characteristics, we find that the current depends on the voltage through a power-law. We find that the magnetic field changes the prefactor of the power-law, whereas the exponent remains unaffected. We also studied the effect of the magnetic field on the electroluminescence (MEL) of the devices and analyze the relationship between MR and MEL. We find that the largest part of MEL is simply a consequence of a change in device current caused by the MR effect.

\end{abstract}

%Uncomment for PACS numbers title message

% Uncomment for Submitted to journal title message

% Comment out if separate title page not required
%\maketitle

\section{Introduction}

Organic $\pi$-conjugated semiconductors (OSEC), which are usually divided into the classes of small molecular weight compounds and macromolecular polymers, respectively have been used to manufacture promising devices such as organic light-emitting diodes (OLEDs) \cite{ElectroluminescenceReview,Alq3Electroluminescence}, photovoltaic cells \cite{PVCellReview,Peumans:2003} and field-effect transistors \cite{Sirringhaus:1999,reviewFET}. A conjugated polymer is a carbon-based macromolecule through which the valence $\pi$-electrons are delocalized. Research into the electronic and optical properties of conjugated polymers began in the 1970's after a number of seminal experimental achievements. First, the synthesis of polyacetylene thin films \cite{Ito:1974} and the subsequent success in doping these polymers to create conducting polymers \cite{Chiang:1977} established the field of synthetic metals. Second, the synthesis of phenyl-based polymers (e.g. poly(para-phenylene vinylene) and discovery of electroluminescence (EL) under low voltages in these systems \cite{Burroughes:1990} established the field of polymer optoelectronics.

In addition to $\pi$-conjugated polymers, small molecular-weight, organic compounds have also been extensively investigated. EL from OLEDs made from small molecules was first observed and extensively studied in the 1960s \cite{AnthraceneElectroluminescence1}. In 1987, a team at Kodak introduced a double layer OLED, which combined modern thin film deposition techniques with suitable materials and structure to give moderately low bias voltages and attractive luminance efficiency \cite{Alq3Electroluminescence}. Intense research in both academia and industry has yielded OLEDs with remarkable color fidelity, device efficiency, and operational stability. In particular, tris-(8-hydroxyquinoline) aluminum (Alq$_3$) has emerged as a widely used electron-transporting and light-emitting material in OLEDs \cite{Hung:2002}.

Recently there has been growing interest in spin \cite{nature,TalianiPaper,ValyNature} and magnetic field effects \cite{Bussmann,Preprint,Bussmann2} in organic semiconducting materials. Xiong et al. \cite{ValyNature} recently demonstrated the first organic semiconductor spin-valve based on Alq$_3$; Davis and Bussmann \cite{Bussmann2} showed that the electroluminescence intensity can be modulated in OLEDs based on the same small molecule by application of a magnetic field. While studying semiconducting polymer OLEDs made from polyfluorene (PFO) we surprisingly discovered \cite{Francis:2004} a large and intriguing magnetoresistance (MR) effect, which we dubbed Organic Magnetoresistance (OMAR). In our best polyfluorene devices the OMAR effect reached up to 10\% (defined as $\Delta R/R \equiv (R(B)-R(0))/R(0)$; R is the device resistance) at room temperature for magnetic fields, B=10 mT. The OMAR effect is therefore amongst the largest of any bulk material. The polymer devices we described can be manufactured cheaply on flexible substrates, and can be transparent. Our devices therefore hold promise for applications where large numbers of MR devices are needed, such as magnetic random-access-memory (MRAM); and applications related to OLED display screens such as touch screens where the position of a magnetic stylus is detected. Our devices do not require ferromagnetic electrode materials resulting in a flexibility in material choice not achievable for other MR devices.

In Ref. \cite{Francis:2004} we show that the OMAR effect in polyfluorene is largely independent of the electron-injection material and therefore is related to the hole-transport through the polymer film. In addition, we found that OMAR increases with lowering the barrier for injection of holes. This shows that OMAR is not related to an interface resistance effect. In addition we found that the OMAR effect is largely independent of film thickness, the onset voltage of the devices was however proportional to the thickness.

Having demonstrated OMAR in polymers, it is natural to ask whether OMAR also exists in small molecules. This extension would be highly relevant both from the application as well as the scientific point of view. Whereas polymers are quasi-one-dimensional, Alq$_3$ corresponds more to quasi-zero-dimensional. Whereas polyfluorene and most other $\pi$-conjugated polymers are hole-conductors, meaning that the hole mobility greatly exceeds that for electrons \cite{Redecker:1998}, Alq$_3$ is an electron transporter. In addition, in polymers it was found that the interaction cross sections between electrons and holes are spin-dependent \cite{nature,Friendnature}, whereas they were found to be spin-independent in Alq$_3$ \cite{Baldo,rAlq3}. Therefore it is non-trivial that OMAR would occur in Alq$_3$ even if it occurs in polymers. Nevertheless, here we report on the observation of a large OMAR effect in Alq$_3$ devices.

In the following we will describe the device fabrication, the MR measurements, and perform an extensive characterization of the OMAR effect in Alq$_3$ devices. We anticipate that a theoretical understanding of this OMAR effect will lead to advances in the understanding of transport processes in organic semiconductors.

\section{Experimental}

\begin{figure}
\begin{center}
\epsfxsize=3.25in
\epsfbox{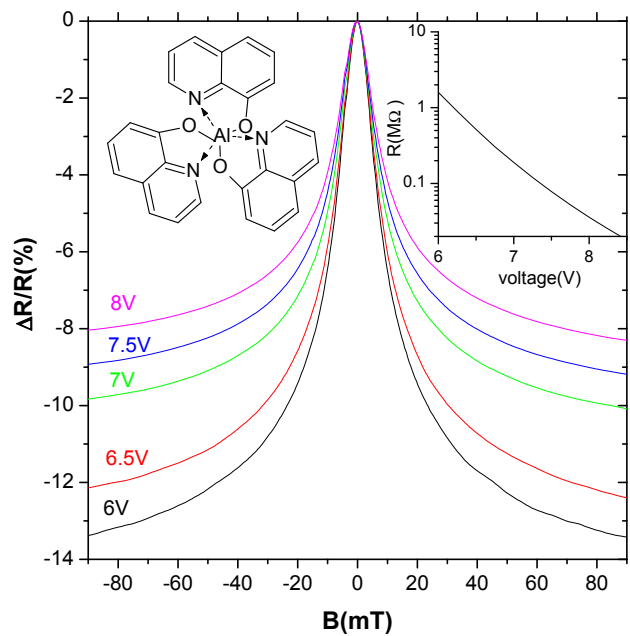}
\end{center}
\caption{\label{fig:Fig1} Magnetoresistance, $\Delta R/R$ curves, measured at room temperature in an ITO (30 nm)/PEDOT ($\approx$ 100nm)/Alq$_3$ ($\approx$ 50 nm)/Ca ($\approx$ 50nm including capping layer) device at different voltages. The inset shows the device resistance as a function of the applied voltage.}
\end{figure}

Our thin film sandwich devices consist of the small molecule Alq$_3$ (see Fig. 1 inset) sandwiched between a top and bottom electrode. Alq$_3$ was purchased from H. W. Sands corp. and was used as received. The Alq$_3$ film was fabricated by evaporation at a base pressure of $10^{-6}$ mbar. The bottom electrode consisted of either indium-tin-oxide (ITO) covered glass or doped poly(3,4-ethylenedioxythiophene) poly(styrenesulfonate) (PEDOT) spin-coated on top of ITO. The top contact, either Al, Ca (covered by a capping layer of Al), or Au, was evaporated through a shadow mask (active area: 1 $mm^2$) at a base pressure of $10^{-6}$ mbar. All manufacturing steps were performed inside a nitrogen glove-box. The MR two-terminal measurements were performed with the sample mounted on the cold finger of a closed-cycle He cryostat located between the poles of an electromagnet. The MR was determined by measuring the current at a constant applied voltage, V. The magnetic field effect on the EL (MEL) was measured at either constant device voltage or constant current using a photomultiplier (PMT) tube located $\approx$ 5 cm outside the magnet poles in order to minimize effects of B on the PMT electron current.

\section{Results and Discussion}

Fig.~\ref{fig:Fig1} shows measured OMAR traces in an Alq$_3$ sandwich device (details are given in the caption) at room-temperature at different Vs. In agreement with our previous results in PFO, we found that the measured MR traces in Alq$_3$ devices are independent of the angle between film plane and applied magnetic field. All measurements shown were performed with an in-plane magnetic field.

\subsection{OMAR devices using different electrode materials}

\begin{figure}
\begin{center}
\epsfxsize=3.25in
\epsfbox{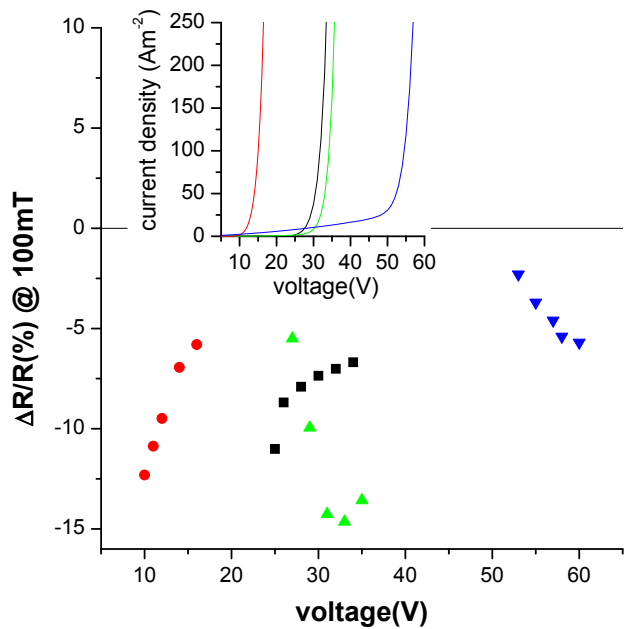}
\end{center}
\caption{\label{fig:Fig2} Dependence of $\Delta R/R$ at 100 mT and 300 K on the device voltage in a variety of Alq$_3$ devices with different electrode materials. The inset shows the current-voltage characteristics of these devices using the same color code as for $\Delta R/R$. $\textcolor[rgb]{1.00,0.00,0.00}{\bullet}$ is for PEDOT/Alq$_3$ ($\approx$ 150 nm)/Ca, $\textcolor[rgb]{0.00,0.00,0.00}{\blacksquare}$ is for ITO/Alq$_3$ ($\approx$ 150 nm)/Ca, $\textcolor[rgb]{0.00,1,0.00}{\blacktriangle}$ is for PEDOT/Alq$_3$ ($\approx$ 150 nm)/Al, $\textcolor[rgb]{0.00,0.00,1.00}{\blacktriangledown}$ is for PEDOT/Alq$_3$ ($\approx$ 150 nm)/Au.}
\end{figure}

Before we discuss the electrode dependence of OMAR in Alq$_3$ devices, let us recall our previous results \cite{Francis:2004} in PFO devices. We found that in PFO both OMAR and current-voltage (I-V) characteristics were largely independent of the electron injecting material. In addition, it was found that lowering the barrier for hole injection results in a reduced onset voltage and an increase in OMAR magnitude. This shows that OMAR in PFO is related to the hole transport. The electron mobility in Alq$_3$ is about 100 times larger than the hole mobility \cite{Kepler:1995} (this is in grave distinction to PFO, where the hole mobility greatly exceeds that of electrons \cite{Redecker:1998}). We would therefore expect that the hole-injecting electrode should have little influence on either the I-V or OMAR response of the Alq$_3$ device. A change in electron injector, e.g. going from a lower to a higher workfunction metal, on the other hand, should result in an increase in onset voltage and a decrease in OMAR magnitude due to a large electron-injecting interface resistance.

Fig.~\ref{fig:Fig2} shows the dependence of the magnitude of the OMAR effect at 100 mT and 300 K on V in a variety of Alq$_3$ devices using different electrode materials (details are given in the caption). PEDOT and Ca are commonly used in OLEDs since they result in relatively small barriers for hole and electron injection, respectively. ITO is another common contact for hole injection because of its large work function. We used Ca, Al, or Au as the top electrode material, resulting in efficient (Ca), moderately efficient (Al) and inefficient (Au) electron injection. The current-voltage (I-V) characteristics of the measured devices are shown as an inset to Fig.~\ref{fig:Fig2}. It is seen that the I-V curves are strongly non-linear as is usually the case in organic sandwich devices. We found that both I-V and OMAR response critically depend on the choice of electron-injecting cathode material choice. Ca cathodes result in low onset voltage and large OMAR response, whereas using Al results in a drastic increase in onset V and decrease in OMAR magnitude at small currents. At high voltages, however the OMAR response becomes as large as in Ca cathode devices, presumably because the large cathode interface barrier is overcome at large V. The situation is even more drastic when using a Au cathode. This observation can easily be rationalized since Alq$_3$ is an electron transporter and Ca has the lowest work-function, followed by Al, whereas Au is known to be unsuitable for electron injection. The increased onset voltage and decreased OMAR can be rationalized considering the increase in the electron-injection barrier and the resulting increase of the interface series resistance, respectively when using high work function anodes. Since the hole mobility is $\approx$ 100 times smaller than the electron mobility \cite{Kepler:1995}, we expect that the I-V and OMAR responses should be largely independent of the choice of the anode material. This is indeed the case regarding the magnitude of the OMAR response at comparable device currents (comparing the PEDOT/Alq$_3$/Ca device with the ITO/Alq$_3$/Ca device), but we clearly observe a considerable change in onset voltage when changing the anode material from PEDOT to ITO. It therefore appears that holes also play a role, at least in determining the onset voltage. The ultimate test would be to manufacture electron only devices, consisting of Ca anode and cathode. However, in fabricating such devices we encountered problems with oxidation of the bottom Ca electrode which we could not overcome. Most of the data shown in the remaining part of the manuscript is measured in PEDOT/Alq$_3$/Ca devices since this is the preferred OLED configuration. No MR effect was observed in ITO/PEDOT/Ca devices.

\subsection{MR devices using different Alq$_3$ film thickness}

\begin{figure}
\begin{center}
\epsfxsize=3.25in
\epsfbox{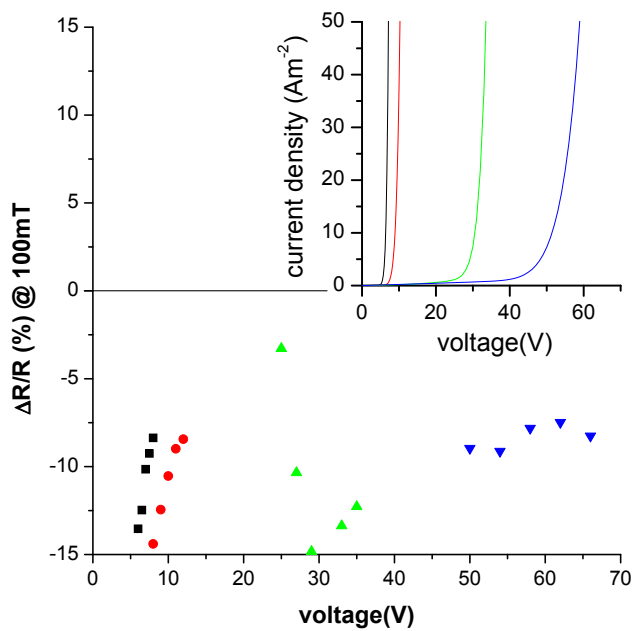}
\end{center}
\caption{\label{fig:Fig3} Dependence of $\Delta R/R$ at 100 mT and 300 K on the device voltage in a variety of devices with different Alq$_3$ film thickness. The inset shows the I-V characteristics of these devices using the same color code as for $\Delta R/R$. $\blacksquare$ is for an ITO/Alq$_3$ ($\approx$ 50 nm)/Ca device, $\textcolor[rgb]{1.00,0.00,0.00}{\bullet}$ is for ITO/Alq$_3$ ($\approx$ 100 nm)/Ca, $\textcolor[rgb]{0.00,1.00,0.00}{\blacktriangle}$ is for ITO/Alq$_3$ ($\approx$ 200 nm)/Ca, and $\textcolor[rgb]{0.00,0.00,1.00}{\blacktriangledown}$ is for ITO/Alq$_3$ ($\approx$ 400 nm)/Ca.}
\end{figure}

Fig.~\ref{fig:Fig3} shows the dependence of the magnitude of the OMAR effect in ITO/Alq$_3$/Ca devices with different polymer film thickness (details are given in the caption) on V. We found that the onset voltage in the linear-linear I-V plot in these devices is approximately proportional to the Alq$_3$ film thickness. In Fig.~\ref{fig:Fig3} it is seen that $\Delta R/R$ typically increases in magnitude with increasing R. However, we find that R of our devices decreases much faster with increasing V than does the magnitude of the MR effect. This suggests that the "intrinsic" MR may be entirely independent of R. The actually observed weak dependence of $\Delta R/R$ on R may be related to series resistances outside of the PFO film, such as hole-injection (Schottky-like) interface resistance. This idea is supported by the observation that $\Delta R/R$ becomes more and more voltage independent as the film thickness increases and thereby the influence of the interface resistance decreases.

\subsection{Temperature dependence}

\begin{figure}
\begin{center}
\epsfxsize=5.5in
\epsfbox{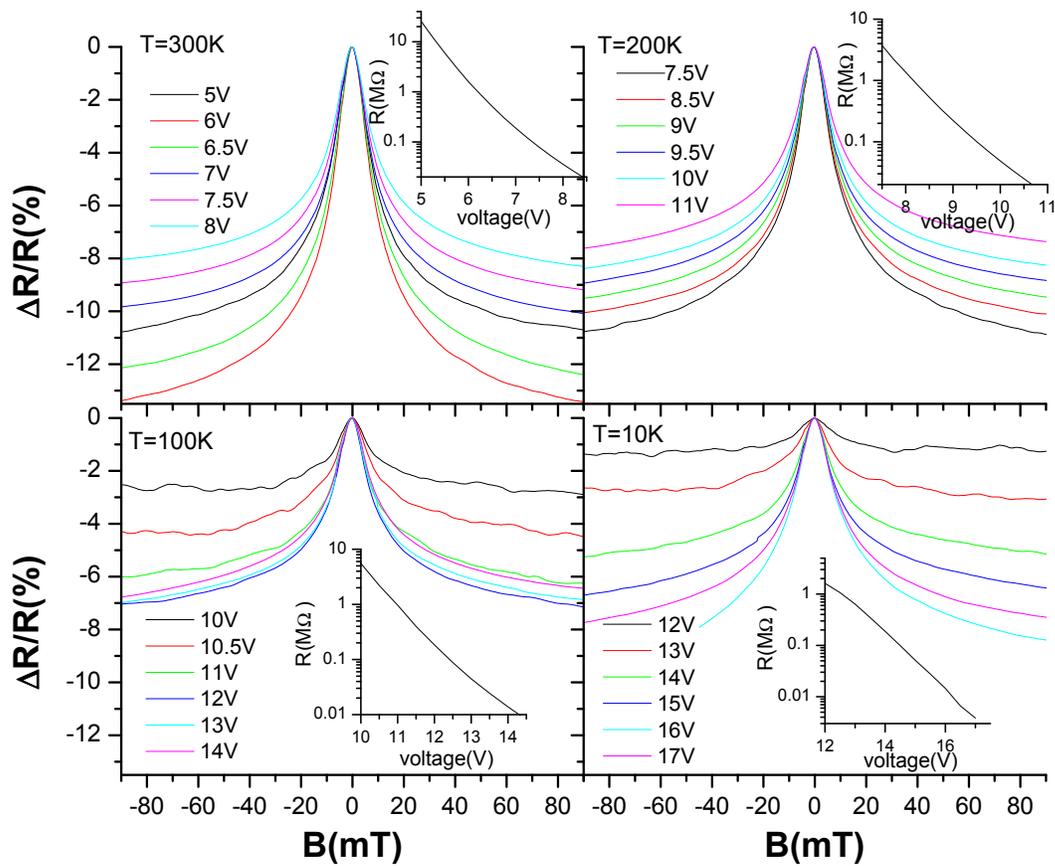}
\end{center}
\caption{\label{fig:Fig4} Magnetoresistance, $\Delta R/R$ curves in the device of Fig.~\ref{fig:Fig1} measured at different temperatures, namely 10 K, 100 K, 200 K, and 300 K. The applied voltages are assigned. The insets show the device resistance as a function of the applied voltage.}
\end{figure}

Fig.~\ref{fig:Fig4} shows MR traces in a PEDOT/Alq$_3$/Ca device for four different temperatures between 300 K and 10 K. We find that the magnitude and width of the MR cones are relatively insensitive to temperature. Fig.~\ref{fig:Fig4}, inset shows R as a function of V at the different temperatures.

\subsection{Relation between magnetic field effects on resistance, current, voltage and electroluminescence}

\begin{figure}
\begin{center}
\epsfxsize=4in
\epsfbox{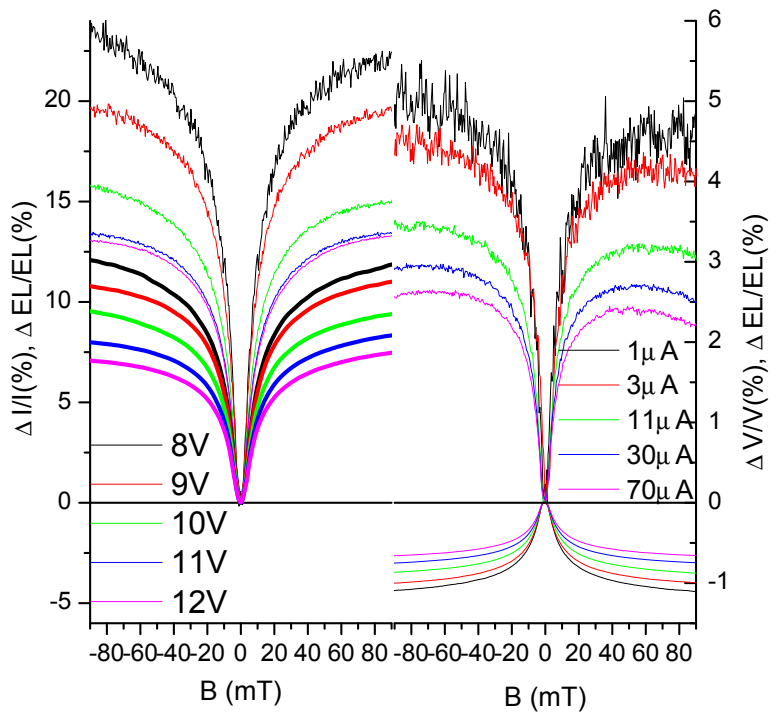}
\end{center}
\caption{\label{fig:Fig5} Left panel: Magnetocurrent, $\Delta I/I$ (bold lines) curves and magnetoluminescence, $\Delta EL/EL$ (thin lines) measured at constant V in a PEDOT/Alq$_3$ (150nm)/Ca device at 300 K. The applied voltages are assigned. Right panel: Magnetovoltage, $\Delta V/V < 0$ curves and magnetoluminescence, $\Delta EL/EL > 0$ (thin lines) measured at constant I in a PEDOT/Alq$_3$ (150nm)/Ca device at 300 K. The device currents are assigned. The currents for the curves in the right panel were chosen such that they approximately coincide between curves of the same color in the left and right panels.}
\end{figure}

Thus far we have discussed our results for $\frac{\Delta I}{I}|_V$ (this notation means that we measured a change $\Delta I$ in current, I caused by B while keeping V constant) that we plotted as $\Delta R/R$. We may also study $\frac{\Delta V}{V}|_I$, i.e. a change $\Delta V$ in V caused by B while keeping I constant. The results, measured in a PEDOT/Alq$_3$/Ca device, for $\frac{\Delta I}{I}|_V$ are shown in Fig.~\ref{fig:Fig5} in the left panel, whereas $\frac{\Delta V}{V}|_I$ is shown in the right panel.

\subsubsection{Magnetocurrent and magnetovoltage}

I and V are related through

\begin{eqnarray}
% \nonumber to remove numbering (before each equation)
  I &=& I_e + I_h \label{eq:IV1} \\
  & \approx & I_e \label{eq:IV2} \\
  &=& \frac{S}{d} \times n e \mu V \label{eq:IV3} \\
  & \approx & A V^\alpha \label{eq:IV4} 
\end{eqnarray}  

\begin{figure}
\begin{center}
\epsfxsize=4in
\epsfbox{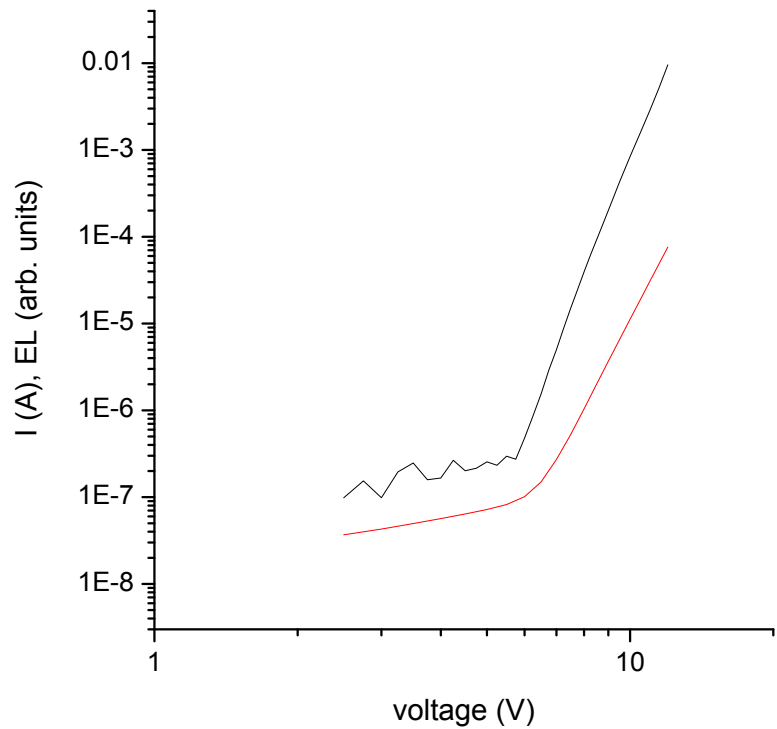}
\end{center}
\caption{\label{fig:Fig6} Current-voltage (red) and EL-voltage (black) characteristics for the device of Fig.~\ref{fig:Fig5}.}
\end{figure}

Eq.~\ref{eq:IV1} expresses that both electron, $I_e$ and hole currents, $I_h$ are present in PEDOT/Alq$_3$/Ca devices. Since the electron mobility, $\mu$ is roughly 100 times larger than the hole mobility \cite{Kepler:1995}, we expect that the current is mostly carried by electrons (Eq.~\ref{eq:IV2}). Eq.~\ref{eq:IV3} gives the relation for $I_e$ in terms of (mobile) electron density, n and mobility, $\mu$; device area, S and film thickness, d; e is the elementary charge. We note that in addition to mobile electrons, a considerable density of immobile, i.e. trapped electrons may also be present which together with n forms the total negative space charge. The relationship Eq.~\ref{eq:IV4} follows from the experimental data shown in Fig.~\ref{fig:Fig6} and is valid for I $> .5 \mu$A. We find $\alpha \approx 10.5$. We note that a power-law relationship between I and V is commonly observed in OLEDs and is usually interpreted using a model of space-charge limited current in the presence of traps \cite{Blom:2000}:

\begin{eqnarray}
% \nonumber to remove numbering (before each equation)
  A &=& N_C e \mu \left (\frac{\epsilon_0 \epsilon_r}{e N_t} \right )^r \frac{1}{d^{2r+1}} C(r) \\
  \alpha &=& r+1 \\
  r &=& \frac{E_t}{kT} \\
  C(r) &=& r^r(2r+1)^{r+1}(r+1)^{-r-2}
\end{eqnarray}  

where $N_C$ is the effective conduction band density of states, $N_t$ is the trap density and $E_t$ the characteristic trap energy, $\epsilon_0$ and $\epsilon_r$ are the vacuum and relative permeability, respectively, k is the Boltzmann constant and T the temperature. There are two distinct possibilities for the magnetic field effect on the current $I=I(B)$, namely (i) $A=A(B)$ or (ii) $\alpha=\alpha(B)$. From relationship Eq.~\ref{eq:IV4} we can calculate

\begin{figure}
\begin{center}
\epsfxsize=4in
\epsfbox{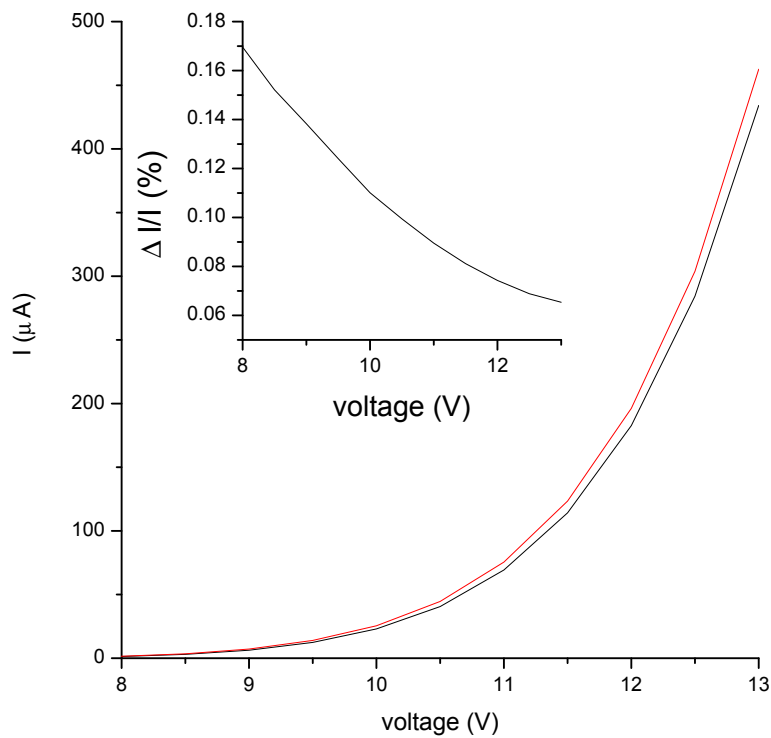}
\end{center}
\caption{\label{fig:IVB} Current-voltage (I-V) characteristics with (red, B = 100 mT) and without magnetic field (black) in a PEDOT/Alq$_3$ (100nm)/Ca device. The inset shows the magnetocurrent, $\Delta I/I$ as directly calculated from the I-V curves.}
\end{figure}

\begin{eqnarray}
    (i) \frac{\Delta I}{I}|_V = \frac{\Delta A}{A} \label{eq:DII1} \\
    (ii) \frac{\Delta I}{I}|_V = \Delta \alpha \label{eq:DII2}
\end{eqnarray}

Since it is difficult to reliably determine small differences, $\Delta \alpha$ of order 0.1 in $\alpha \approx 10.5$ from the I-V characteristics (see Fig.~\ref{fig:IVB}, we will now use an indirect but more reliable method to distinguish between scenarios (i) and (ii): From relationship Eq.~\ref{eq:IV4}, we can also calculate

\begin{eqnarray}
    (i) \frac{\Delta V}{V}|_I & = & - \alpha^{-1} \frac{\Delta A}{A} = - \alpha^{-1} \frac{\Delta I}{I}| _V \label{eq:DVV1} \\
    (ii) \frac{\Delta V}{V}|_I & = & - ln(V) \frac{\Delta \alpha}{\alpha} = - ln(V) \alpha^{-1} \frac{\Delta I}{I}| _V \label{eq:DVV2}
\end{eqnarray}

The right most relations in the above equations were obtained by combining the left most equations of Eqs.~\ref{eq:DVV1} and \ref{eq:DVV2} with Eqs.~\ref{eq:DII1} and \ref{eq:DII2}. Experimentally we find (see Fig.~\ref{fig:Fig5}) $\frac{\Delta V}{V}|_I = -11^{-1} \frac{\Delta I}{I}|_V$ in good agreement with expectation based on scenario (i), but in disagreement with scenario (ii). We therefore conclude $I(B)=A(B)V^\alpha$.

\subsubsection{Magnetoluminescence}

EL and I are related through

\begin{eqnarray}
% \nonumber to remove numbering (before each equation)
  EL & \propto & I \eta_{EL} \label{eq:EL1} \\
  & = & b I^\beta \label{eq:EL2} \\
  \Rightarrow \eta_{EL} & \propto & I^{\beta-1} \label{eq:EL3}
\end{eqnarray}  

$\eta_{EL}$ is the EL quantum efficiency defined as the ratio of the number of emitted photons (which is proportional to the measured EL intensity) to the number of carriers flowing in the external circuit (which is proportional to the device current). The relationship Eq.~\ref{eq:EL2} follows from the experimental data shown in Fig.~\ref{fig:Fig6} and is valid for I $> .2 \mu$A. We find $\beta \approx 1.5$ for $0.2 \mu A < I < 10 \mu A$ and $\beta \approx 1.3$ for $10 \mu A < I$. From relationship Eq.~\ref{eq:IV4} between EL and I, we can calculate, assuming that B influences the current prefactor only,

\begin{equation}\label{eq:DVV}
    \frac{\Delta EL}{EL}|_V = \beta \frac{\Delta I}{I}|_V
\end{equation}

Experimentally we find (see Fig.~\ref{fig:Fig5}) $\frac{\Delta EL}{EL}|_V = 1.8 \frac{\Delta I}{I}|_V$ for I $= 3 \mu$A and $\frac{\Delta EL}{EL}|_V = 1.65 \frac{\Delta I}{I}|_V$ for I $= 30 \mu$A in reasonable agreement with expectation (i.e. ratios of 1.5 and 1.3, respectively). However, the $\approx 25\%$ discrepancy between predicted and measured ratios indicates that the magnitude of I does not uniquely determine EL. This is not unexpected since it is known that $\eta_{EL}$ in Eq.~\ref{eq:EL1} also depends e.g. on the balance between electron and hole carriers, and on the position of the recombination zone, both of which could shift as a result of the magnetic field (effect on the current). We may directly measure the magnetic field effect on $\eta_{EL}$ at constant current (i.e. any effects in addition to the dependence of $\eta_{EL}$ on the \emph{magnitude} of I) by measuring $\frac{\Delta EL}{EL}|_I$. The experimental traces for the magnetic field effect on the EL at constant current, $\frac{\Delta EL}{EL}|_I$ are shown in Fig.~\ref{fig:Fig5} right panel. We find that $\frac{\Delta EL}{EL}|_I \ll \frac{\Delta EL}{EL}|_V$. This indicates that most of the magnetoluminescence effect is simply due to the fact that the magnitude of the current changes (indeed, we find with good accuracy that $\frac{\Delta EL}{EL}|_V=\beta \frac{\Delta I}{I}|_V + \frac{\Delta EL}{EL}|_I$). This in turn indicates that the magnetotransport effect is the primary effect, whereas the magnetoluminescence effect is secondary, i.e. a consequence of the magnetotransport effect. This conclusion is in agreement with our previous results in PFO where we showed that OMAR exists also in hole-only devices.

\subsection{Universality of OMAR traces in polymers and small molecules}

\begin{figure}
\begin{center}
\epsfxsize=4in
\epsfbox{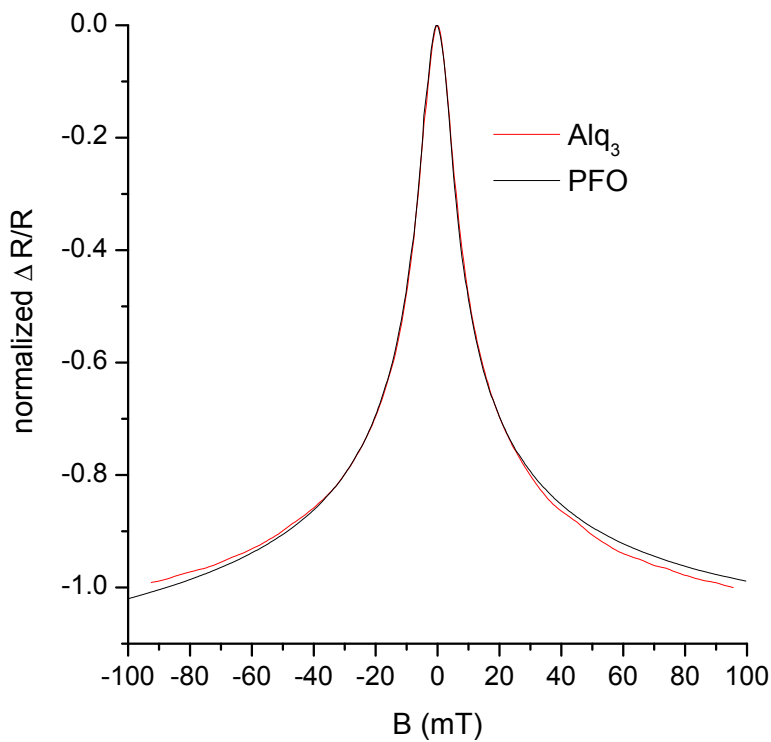}
\end{center}
\caption{\label{fig:Universality} Normalized magnetoresistance, $\Delta R/R$ curve of the device of Fig.~\ref{fig:Fig1} and that of a PEDOT/PFO ($\approx$ 100 nm)/Ca device measured at 300 K.}
\end{figure}

Fig.~\ref{fig:Universality} shows the normalized $\Delta R/R$ traces in PEDOT/$Alq_3$/Ca and PEDOT/PFO/Ca devices. It is seen that the functional dependence is exactly identical in both devices. This is very surprising, since the chemical structures of the two materials are quite different, and one therefore expects that they possess quite different material parameters such as transport properties. The "universality" of the OMAR traces therefore implies that the explanation of the OMAR effect must be quite general and simple (in the sense that detailed material properties cannot enter).

\section{Summary}

In summary, we discovered a large MR effect in Alq$_3$ sandwich devices that is similar to the OMAR effect previously discovered in polymer devices. The magnitude of the negative MR effect is several percent at fields on the order of 10mT dependent on V. The effect is independent of the sign and direction of the magnetic field, and is only weakly temperature dependent. Both electron and hole currents appear to participate in the OMAR effect in Alq$_3$. We find that the magnetic field changes the prefactor of the power-law relating the current to the voltage, whereas the exponent remains largely unaffected. We also studied the effect of the magnetic field on the electroluminescence (MEL) of the devices and analyzed the relationship between MR and MEL. We find that the largest part of MEL is simply a consequence of a change in device current caused by the MR effect. This suggests that the magnetotransport effect is the primary effect, whereas the magnetoluminescence is a consequence of the magnetotransport. This is in agreement with our finding in hole-only PFO devices, which show large magnetoresistance even though there is no EL.

\ack We acknowledge fruitful discussions with Profs. M. E. Flatt\'{e} and Z. V. Vardeny. This work was supported by Carver foundation and NSF ECS 04-23911.

\newpage
\bibliography{magnetoresistance}

\end{document}